\documentclass[proceedings, preprint]{rmaa}

\usepackage{paralist}

\usepackage{psfrag,color}

\newcommand{\Ha}{H$\alpha$}
\newcommand{\Hb}{\ifmmode {\rm H}\beta \else H$\beta$\fi}

\newcommand{\hii}{H~{\sc ii}}

\newcommand{\nii}{[N~{\sc ii}]}

\newcommand{\Oiii}{[O~{\sc iii}] $\lambda$5007}

\SetYear{2011}
\SetConfTitle{LARIM XIII}

\title{HOLMES and little monsters} 

\author{
  G. Stasi\'nska\altaffilmark{1}}

\altaffiltext{1}{LUTH, Observatoire de Paris, CNRS, Universit\'e Paris
  Diderot; Place Jules Janssen 92190 Meudon, France\\(grazyna.stasinska@obspm.fr).}

\shortauthor{Stasi\'nska}
\shorttitle{HOLMES and little monsters}

\listofauthors{G. Stasi\'nska}
\indexauthor{Stasi\'nska, G.}

\abstract{Hot low-mass evolved stars (HOLMES) may have a more important role in  galaxies than generally thought.}

\resumen{Estrellas evolucionadas calientes de baja masa podrian tener un papel mas importante de lo generalmente pensado.}

\addkeyword{galaxies: active}
\addkeyword{galaxies: elliptical and lenticular, cD}
\addkeyword{galaxies: ISM}
\addkeyword{stars: AGB and post-AGB}

\begin{document}
\maketitle

\section{Present status}
\label{sec:intro}

There are several families of emission-line objects in the Universe whose source of ionization is still debated.

\subsection{LINERS and alike}

 The most known is the case of LINERs (low-ionization nuclear emission regions), defined as a class by Heckman (1980) and which are found in nearly 40\% of nearby galaxies (Ho 2008). The spectra of these objects are dominated by lines from low ionization states. This distinguishes them both from Seyfert galaxies and from \hii\ regions. In addition, the line luminosities are modest. While the first explanations for the observed emission-line spectra favoured the effects of shock waves, it now seems that many of these nuclei harbour a supermassive black hole accreting at a low rate,  a ``little monster'' as called by Eracleous et al. (2010). While this interpretation seems to hold for an important fraction of LINERs (Gonz{\'a}lez-Mart{\'{\i}}n et al. 2009), other possibilities have to be investigated for those objects that apparently are not ``little monsters''. Ionization by  hot low-mass evolved stars (HOLMES) is one of them. Such an explanation was already proposed  by Binette et al. (1994) to account  for the emission lines observed in early-type galaxies.

With the advent of the Sloan Digital Sky Survey (SDSS, York et al. 2000) which provided  fiber-fed spectra of a million of galaxies, it was found that a large fraction of galaxies have spectral characteristics of LINERs (Kauffmann et al. 2003, Kewley et al. 2006). The typical redshift of those galaxies is 0.1, so that the  3\arcsec\ fibers cover regions much larger than just the nuclei. Using the spectral synthesis code {\sc STARLIGHT} (Cid Fernandes et al. 2005),  Stasi\'nska et al. (2008) showed that the stellar populations of those galaxies are old, and that the Lyman continuum photons produced by these old stellar populations are able to explain the location of these ``LINER galaxies'' in the classical emission-line ratio diagram of Phlillips, Baldwin \& Terlevich, 1981, BPT): indeed, the integrated spectra of the HOLMES in these old stellar populations are much harder than those of the massive stars that ionize pure star-forming galaxies. 

Cid Fernandes et al. (2010) studied the  ``forgotten'' population of emission-line galaxies in the SDSS, i.e. galaxies that do not appear in the BPT diagram because their \Oiii\ and/or \Hb\ lines are too weak. By introducing the  W(\Ha) vs. \nii/\Ha\ (WHAN) diagram they could classify 50\% more emission-line galaxies. It turns out that a large fraction of them are LINER-like galaxies, defined by  log \nii/\Ha\ $> -0.4$ and W(\Ha) $<6$\AA. 

The question of how to distinguish ``retired'' galaxies, ie., galaxies that have stopped forming stars and are ionized by their old stellar population from galaxies ionized by ``little monsters'', or, in other words, fake from true AGN, was discussed in Cid Fernandes et al. (2011). A practical limit between these two overlapping populations of LINER-like galaxies was set to W(\Ha) $=3$\AA. It was corroborated by the analysis of global properties of the different spectral classes of galaxies defined in the WHAN diagram, showing similarities between galaxies hosting strong or weak AGN on the one hand, and galaxies which have stopped forming stars, i.e. ``retired'' and ``passive'' (or line-less),  on the other.

\subsection{The diffuse ionized gas in late-type galaxies}

Another case where the cause of the  ionization is disputed is that of the diffuse ionized gas (DIG) in spiral galaxies. As shown by Flores-Fajardo et al. (2009),  DIG spectra fall  in the LINER zone of the BPT diagram. This means that heating in the DIG is more efficient than heating by massive stars in \textit{classical} \hii\ regions. Haffner et al. (2009) give a review of the possible heating sources: ionizing photons from leaking \hii\ regions,   shocks, dissipation of turbulence, magnetic reconnection etc... They also mention the possible role of pre-white dwarfs. Flores-Fajardo et al. (2011 and these proceedings) have shown that,  the \textit{expected} content of OB stars and HOLMES in the well observed edge-on spiral galaxy NGC 891 can well explain the observed emission-line characteristics  of the extraplanar DIG in this galaxy. This galaxy was chosen because it offered the most complete set of observational constraints that should be fitted by the model and not because it is a special case. It is likely that HOLMES (complementing the radiation from OB stars in leaky \hii\ regions.) are indeed the right answer for the ionization of the DIG in all spiral galaxies. 

\section{The future}
\label{sec:future}

So HOLMES provide a good and natural explanation to the problem of heating and ionization of gas in various galactic contexts.  We know that HOLMES exist, and in large amounts. They provide sufficient numbers of ionizing photons to produce detectable emission lines. The high energies of these photons (roughly similar to those of a $10^5$\,K star), are able to yield LINER-like emission-line spectra. Of course, the ionizing radiation field from HOLMES can be easily dominated by other sources (massive stars or accreting black hole). But they provide a floor in W(\Ha)  below which no galaxy can be found (at least if it contains warm gas). 

So far,  studies of HOLMES in galaxy context have been scarce. A few recent works use integral field spectroscopy to distinguish between HOLMES and ``little monsters'' as an ionization source in early-type galaxies (Annibali et al. 2010, Sarzi et al. 2010). Progress on the theoretical side requires 
models for the complete evolution of low- and intermediate-mass stars at all metallicities, as well as synthetic stellar populations based on them and using relevant atmosphere models with appropriate chemical composition. The physics of advanced stages of stellar evolution is complex but  progress is being made (Marigo et al. 2008, Weiss \& Ferguson 2009). Another   issue is the physical properties of the interstellar matter in early-type galaxies. The gas is believed to be essentially material lost trough winds from low- and intermediate mass stars, and thus should be enriched in carbon, nitrogen and dust. But  its  distribution within the galaxies is not known. Also, it is not clear why certain galaxies that stopped forming stars are line-less and others are not. One possibility is that gas may have been swept out of the galaxies by some dynamical process, another one is that all the gas remaining in the galaxy has been heated up to coronal temperatures.

The paper by Flores-Fajardo, Morisset, Stasi\'nska \& Binette on the role of hot evolved stars in the DIG, submitted to ApJ in July 2010 was rejected by the editors and will be available elsewhere.
I thank my colleagues from the SEAGAL-STARLIGHT project, in particular R. Cid Fernandes, N. Vale Asari, A. Mateus from UFSC (Brazil) as well as  C. Morisset, N. Flores-Fajardo and L. Binette from UNAM (Mexico) for wonderful collaborations on HOLMES, and the SOC of LARIM 2010 for letting give this presentation.


\begin{thebibliography}

\bibitem{} Annibali F., et al., 2010, A\&A, 519, A40 



\bibitem{} Baldwin J.~A., Phillips M.~M., Terlevich R., 1981, PASP, 93, 5 

\bibitem{}  Binette L., Magris C.~G., Stasi{\'n}ska G., Bruzual A.~G., 1994, A\&A, 292, 13 

\bibitem{}  Cid Fernandes R., Mateus A., Sodr{\'e} L., 
Stasi{\'n}ska G., Gomes J.~M., 2005, MNRAS, 358, 363 


\bibitem{}  Cid Fernandes R., Stasi\'nska G., Mateus A., 
Vale Asari N., 2011, arXiv:1012.4426 


\bibitem{}  Cid Fernandes R., Stasi{\'n}ska G., 
Schlickmann M.~S., Mateus A., Vale Asari N., Schoenell W., Sodr{\'e} L., 
2010, MNRAS, 403, 1036 

\bibitem{}  Eracleous M., Hwang J.~A., Flohic H.~M.~L.~G., 2010, ApJS, 187, 135




\bibitem{}  Flores-Fajardo N., Morisset C., Binette L., 2009, RMxAA, 45, 261 

\bibitem{}  Flores-Fajardo N., Morisset C., Stasi\'nska G., Binette, 2011, submitted

\bibitem{}  Gonz{\'a}lez-Mart{\'{\i}}n O., Masegosa J., M{\'a}rquez I., Guainazzi M., Jim{\'e}nez-Bail{\'o}n E., 2009, A\&A, 506, 1107

\bibitem{}  Haffner L.~M., et al., 2009, RvMP, 81, 969  

\bibitem{}  Heckman T.~M., 1980, A\&A, 87, 152 

\bibitem{}  Ho L.~C., 2008, ARA\&A, 46, 475

\bibitem{}  Kauffmann G., et al., 2003, MNRAS, 346, 
1055 

\bibitem{} 
Kewley L.~J., Groves B., Kauffmann G., Heckman T., 2006, MNRAS, 372, 961 

\bibitem{}  Marigo P., et al., 2008, A\&A, 482, 883 


\bibitem{} 
Sarzi M., et al., 2010, MNRAS, 402, 2187 

\bibitem{} Stasi{\'n}ska G., et al., 2008, MNRAS, 391, L29 



\bibitem{}  Weiss A., Ferguson J.~W., 2009, A\&A, 508, 1343 


\bibitem{} 
York D.~G., et al., 2000, AJ, 120, 1579 


\end{thebibliography}
\end{document}